\documentclass[galaxies,article,accept,moreauthors,LaTeX and dvi2pdf]{Definitions/mdpi} 
\firstpage{1} 
\pubvolume{8}
\issuenum{3}
\articlenumber{59}
\pubyear{2020}
\copyrightyear{2020}
\history{Received: 6 July 2020; Accepted: 31 July 2020; Published: date}

\usepackage{subcaption}
\usepackage[countmax]{subfloat}
\Title{X-ray Flux and Spectral Variability of Blazar H~2356-309}

\Author{Kiran A. Wani and Haritma Gaur *}

\AuthorNames{Kiran A. Wani and Haritma Gaur}

\address[1]{Aryabhatta Research Institute of Observational Sciences (ARIES), Manora Peak, Nainital 263002, India; kiiran.wani2@gmail.com\\}

\corres{\hangafter=1 \hangindent=1.05em \hspace{-0.82em}
Correspondence: harry.gaur31@gmail.com}

\abstract{We present the results of timing and spectral analysis of the blazar H 2356-309 using XMM-Newton observations.
This blazar is observed during 13 June 2005--24 December 2013 in total nine observations. Five of the observations show moderate flux
variability with amplitude 1.7--2.2\%. We search for the intra-day variability timescales in these five light curves, but did not find  in any of them.
The fractional variability amplitude is generally lower in the soft bands than in the hard bands, which is attributed to the energy dependent synchrotron emission.
Using the hardness ratio analysis, we search for the X-ray spectral variability along with flux variability in this source. However, we~did not find any significant
spectral variability on intra-day timescales. We also investigate the X-ray spectral curvature of blazar H  2356-309 and found that six of our
observations are well described by the log parabolic model with $\alpha$ = 1.99--2.15 and $\beta$ = 0.03--0.18. Three of our observations are well described
by power law model. The break energy of the X-ray spectra varies between 1.97--2.31 keV. We investigate the correlation between various parameters that are derived from log parabolic model and their implications are discussed.}
\keyword{BL Lacertae objects; general---BL Lacertae objects; individual; H 2356-309---galaxies; active}

\begin{document}
\section{Introduction} \label{sec:intro}
Blazars are highly luminous AGNs (Active Galactic Nuclei) that emit in all accessible wavelengths ranging from radio to high energy
gamma rays. BL 
Lacertae objects (BL Lacs) and flat spectrum radio quasars (FSRQs) collectively belongs to Blazars. Blazars show very strong continuum with
featurless optical spectrum, which is thought to be due to relativistic jets pointing nearly to our line of sight ($\leq$10$^{\circ}$) \citep{Urry Padovani (1995)}.
Blazar emission is mostly non-thermal in nature, shows variable polarization, and emitted continuous Doppler boosted spectra.

The broadband spectral energy distribution of blazar is characterized by a double peaked structure. The first hump at lower energies is attribued to the synchrotron emission from relativistic electrons in a jet, while the high frequency hump is thought to be produced by inverse Compton scattering from the same electron population with the synchrotron photons (SSC, Synchrotron Self Compton models; e.g., \citep{Marscher (1985)}) or with external ambient photons originated in the BLR (Broad Line region), torus (EC, External Compton models; e.g., \citep{Sikora (1994)}). The peak of the low-energy spectral component is found at X-ray/UV energies for high energy peaked blazars (HBLs/HSPs) and at optical/IR (infrared) energies for low-energy peaked blazars (LBLs/LSPs). The peak of the high energy spectral component for HSPs occurs at GeV/TeV energies, while, for LSPs, it is usually at MeV/GeV energies (\citep{PadovaniGiommi95,Abdo (2010)}).

Blazars are highly variable on different timescales across the whole electromagnetic spectrum~\citep{Ulrich (1997)}. If the variability time is less than a day then it is referred to as intra-day variability (IDV) \citep{Wagner (1995)} and variations over a few days to months is known as short term variability (STV). When the variations range from several months to years or even decades, then it is termed as long term variability (LTV)~\citep{Gupta (2004)}. HSP blazars are brightest in X-ray bands and show strong flux variability over diverse time-scales ranging between minutes to years (e.g., \citep{Sembay (1993),Brinkmann (2005),Zhang (2005),Zhang (2008),Gaur (2010), Kapanadze (2014)}).

The blazar H 2356-309 is hosted by an elliptical galaxy \citep{Schwartz (1989)} located at a redshift z = 0.165 ($\pm$0.002)~\citep{Falomo (1991)}.
H 2356-309 is classified as a high frequency peaked blazar as the X-ray spectrum of this blazar was characterized by a broken power law with a synchrotron peak lying at 1.8~keV using BeppoSAX observations \citep{Costamante (2001),Giommi Colafrancesco (2005)}.
It was discovered in the VHE regime by the H.E.S.S. Cherenkov telescope during June to December 2004 \citep{Aharonian et al. (2006)}.
At X-ray energies, H 2356-309 was initially detected by the UHURU satellite \citep{Forman (1978)} and subsequently by the HEAO-I satellite \citep{Wood (1984)}. It~has been observed with many observing facilities, like BeppoSAX \citep{Costamante (2001)}, NuSTAR \citep{Pandey (2018)}, and Cosmic Origin Spectrograph (COS), on board the Hubble Space Telescope (HST) \citep{Fang (2014)}.

Multi-wavelength observations of H 2356-309 with RXTE Satellite, HESS, ROTSE-IIIc, Nancay radio telescope, XMM-Newton \& ATOM telescope \citep{H.E.S.S. Collaboration et al. (2010)}
suggested that the broadband spectral energy distribution (SED) can be simply fitted with a synchrotron self-Compton (SSC) model with a double-peak structure, in which the peaks in X-ray and VHE are produced via synchrotron radiation and inverse Compton scattering, respectively \citep{H.E.S.S. Collaboration et al. (2010)}. H 2356-309 was observed with
NuSTAR in a time exposure of 21.90 ks on 18 May 2016 \citep{Pandey (2018)}. However, they did not find any significant intra-day variability or spectral variation
in their observations.

In the current study, we present XMM-Newton observations of blazar H 2356-309 held in between June 2005--December 2013. Our aim is to study their flux and spectral variability
in the 0.3--10 keV energy range. Additionally, we fit the spectra of H 2356-309 using various models to study their X-ray spectral curvature and to constrain its break energy.
The observing log of the XMM-Newton data for H 2356-309 blazar is given in Table~\ref{table:Observation log}.

\begin{table}[H]
\caption{Observation Log of XMM-Newton data for  Blazar H 2356-309.}
\label{table:Observation log}
\centering
\begin{tabular}{cccc}
\toprule
\textbf{ Obs. Date}	& \multirow{2}{*}{\textbf{Observation ID}}	& \textbf{Total Elapsed} \boldmath{$^{1}$} & \textbf{Exposure Time} \boldmath{$^{2}$}\\
\textbf{(yyyy-mm-dd)}	& & \textbf{Time (ks)} & \textbf{(ks)} \\
\midrule
2005-06-13 & 0304080501  &  16.75 & 16.52 \\
2005-06-15 & 0304080601  & 17.04 & 16.62\\
2007-06-02 & 0504370701  & 129.59 & 109.93 \\
2012-11-18 & 0693500101   & 118.08 & 80.25\\
2013-12-02 & 0722860101   & 22.19 & 14.86 \\
2013-12-03 & 0722860701   & 63.48 & 44.50\\
2013-12-10 & 0722860201  & 105.48 & 73.05\\
2013-12-12 & 0722860301  & 107.75 & 74.99\\
2013-12-24 & 0722860401  & 99.03 & 69.41\\
\bottomrule
\end{tabular}\\
\begin{tabular}{ccc}
\multicolumn{1}{c}{\small $^{1}$ Full time interval for the exposure; $^{2}$ Weighted live time of CCDs in the extraction region. }
\end{tabular}

\end{table}

\section{XMM-Newton Observations and Data Analysis}

Blazar  H 2356-309 is observed by the European Photon Imaging Camera (EPIC) on board the XMM--Newton satellite \citep{Jansen (2001)}. The EPIC instrument provides
imaging and spectroscopy in the energy range from 0.2 to 15 keV with a good angular resolution (PSF = 6 arcsec FWHM) and a moderate spectral resolution (E/$\Delta$E $\sim$ 20--50).
In this work, we have only considered the EPIC-pn data, as it is most sensitive and less affected by the photon pile-up effects.

Data reduction is performed with the use of XMM-Newton Science Analysis System (SAS)\footnote{\url{https://www.cosmos.esa.int/web/xmm-newton/sas}.} version 18.0.0
for the LC extraction.
The XMM-Newton EPCHAIN pipeline is used to generate the event files. We extracted the high energy \mbox{(10 keV < E < 12
keV)} light curve for the full frame of the exposed CCD in order to identify flaring particle background. We did not find any significant background flares in our observations.
We restrict our analysis to the 0.3--10 keV energy range, as~data below 0.3 keV are considerably contaminated by noise events and data above 10 keV are usually dominated
by background flares. From total nine observation data sets, three frames are in Timing mode and six are in Imaging mode.

In imaging mode, source region is extracted using a circle of 50 arcsec radius centered on the source. Background light curve is obtained from the region that corresponds to
circular annulus centered on the source with inner and outer radius of 62.5 arcsec and 125 arcsec, respectively. In timing mode, the source region is extracted using a rectangular box of 20 pixels along RAWX centered on source verticle strip. The background light curve is obtained from the source free region of 20 pixels along RAWX.
Pile up effects are examined for each observation by using the SAS task EPATPLOT and we found that mostly triple and quadruple events are affected by the pile-up effects; hence, we only extracted the single and double events for our observations. We did not find pile up in our observations.
Finally,~we obtained source LCs for the 0.3–10 keV band (corrected for background flux and given in unit of counts~s$^{-1}$), sampled evenly with a fixed bin size of $\Delta$t = 0.5 ks.
Redistribution matrices and  ancillary response files were produced using the SAS tasks  rmfgen and arfgen. The pn spectra (0.6--10 keV) were created by the SAS tool
XMMSELECT and grouped to have at least 30 counts in each energy bin in order to ensure the validity of $\chi^{2}$ statistics.

\section{Analysis Techniques}\unskip

\subsection{Excess Variance}

Blazars show rapid and strong flux variations on diverse timescales across the EM spectrum. To~quantify the strength of the variability, excess variance $\sigma_{XS}$ \citep{Nandra (1997),Edelson (2002)} and fractional rms variability amplitude $F_{var}$  are often calculated. Excess variance is a measure of source intrinsic variance determined by substracting the variance that arises from measurement errors from the total variance of the observed LC. If a LC consisting of N measured flux values $x_{i}$ with corresponding finite uncertainties $\sigma_{err,\:i}$ arising from measurement errors, then the excess variance is calculated as follows:

\begin{equation}
\sigma^{\:2}_{XS} = S^{2} - \bar{\sigma}^{\:2}_{err}
\end{equation}
where $\bar{\sigma}^{\:2}_{err}$ is the mean square error, defined as

\begin{equation}
\bar{\sigma}^{\:2}_{err} = \dfrac{1}{N} \sum_i\: \sigma^{\:2}_{err,\:i}
\end{equation}
and $S^{2}$ is the sample variance of the LC, as given by

\begin{equation}
S^{2} = \dfrac{1}{N-1} \sum_i  (x_{i} - \bar{x})^{2}
\end{equation}
where $\bar{x}$ is the arithmatic mean of  $x_{i}$

The normalized excess variance is $\sigma^{\:2}_{NXS} = \frac{\sigma^{\:2}_{XS}}{\bar{x}^{\:2}}$  and the fractional rms variability amplitude, $F_{var}$~\citep{Edelson Pike Krolik (1990),Rodriguez-Pascual (1997)}, which is the square root of $\sigma^{\:2}_{NXS}$ is thus

\begin{equation}
F_{var} = \sqrt{\dfrac{S^{2} - \bar{\sigma}^{\:2}_{err}}{\bar{x}^{\:2}}}
\end{equation}

The uncertainty on $F_{var}$ is given by \citep{Vaughan (2003)}.

\begin{equation}
err(F_{var}) = \sqrt{\Biggl(\sqrt{\dfrac{1}{2N}} \: \dfrac{\bar{\sigma}^{\:2}_{err}}{\bar{x}^{\:2} F_{var}}\Biggl)^{2}+\Biggl(\sqrt{\dfrac{\bar{\sigma}^{\:2}_{err}}{N}} \: \dfrac{1}{\bar{x}^{\:2}}\Biggl)^{2}}
\end{equation}

In Table~\ref{table:X-ray variability parameter}, dashes indicate that the sample variances were smaller than the mean square errors, so~that no fractional variance could be claimed. When $F_{var}$ > 3$\:err(F_{var})$, then only we consider a strong evidence for variability to be present.

\begin{table}[H]
\caption{X-ray variability parameter $F_{var}$.}
\label{table:X-ray variability parameter}
\centering
\begin{tabular}{ccccc}
\toprule
\textbf{ Obs. Date}	& \multirow{2}{*}{\textbf{Observation ID}}& \textbf{\textit{F}$_{var}$ (\%)} & 	\textbf{\textit{F}$_{var}$ (\%)} & \textbf{\textit{F}$_{var}$ (\%)}\\
\textbf{(yyyy-mm-dd)} & & \textbf{Soft (0.3--2 keV)}& \textbf{Hard (2--10 keV)}	& \textbf{Total (0.3--10 keV)}\\
\midrule
2005-06-13 & 0304080501 & - & - & -  \\
2005-06-15 & 0304080601 & 0.49 $\pm$ 0.84 & - & 0.42 $\pm$ 0.85 \\
2007-06-02 & 0504370701 & 1.62 $\pm$ 0.12 & 4.82 $\pm$ 0.35  & 1.73 $\pm$ 0.11\\
2012-11-18 & 0693500101 & 2.28 $\pm$ 0.12 & 4.8 $\pm$ 0.30 & 2.69 $\pm$ 0.11\\
2013-12-02 & 0722860101 & 0.67 $\pm$ 0.69 & - & - \\
2013-12-03 & 0722860701 & 0.64 $\pm$ 0.43 & 1.78 $\pm$ 1.16 & 0.72 $\pm$ 0.35\\
2013-12-10 & 0722860201 & 1.63 $\pm$ 0.17 & 1.59 $\pm$ 1.09 & 1.71 $\pm$ 0.16 \\
2013-12-12 & 0722860301 & 1.65 $\pm$ 0.17 & 2.01 $\pm$ 0.85 & 1.77 $\pm$ 0.15 \\
2013-12-24 & 0722860401 & 2.02 $\pm$ 0.15 & 3.12 $\pm$ 0.52 & 2.25 $\pm$ 0.14 \\
\bottomrule
\end{tabular}
\end{table}

\subsection{Hardness Ratio}

To search for spectral changes over a broad X-ray band, hardness ratio (HR) is normally used (e.g.,~\citep{Jin (2006),Sivakoff (2004)}). We extracted LCs in two energy bands, here defining (0.3--2 keV) as the soft band and (2--10 keV) as the hard band. We then computed hardness ratio, which is defined as follows:

\begin{equation}
HR = \dfrac{(H - S)}{(H + S)}
\end{equation}
and the error in \emph{HR} ($\sigma_{HR}$) is calculated, as follows:

\begin{equation}
\sigma_{HR} = \dfrac{2}{(H + S)^{2}} \sqrt{S^{2} \sigma^{\:2}_{H}  + H^{2} \sigma^{\:2}_{S} }
\end{equation}
where \emph{S} and \emph{H} are the net count rates in the soft (0.3--2 keV) and hard (2--10 keV) bands, respectively, while $\sigma_{S}$ and $\sigma_{H}$ are their respective errors.

\subsection{Spectral Analysis}

The XSPEC software package version 12.10.1f is used for spectral fitting. The Galactic absorption $n_{H}$ is fixed to be 1.3 $\times$ $10^{20}$ cm$^{-2}$ \citep{Lockman Savage (1995)} and the Xspec
routine “cflux” is used to obtain unabsorbed flux and its error.

As we know that X-ray spectra of blazars are well described by a single power-law or a broken power law. References \citep{Massaro (2004),Massaro (2008)} found that there is curvature
in the blazar spectra which arise due to log parabolic electron distributions. Hence, they are also well described by the log parabola model (e.g.,~\citep{Tramacere (2007a),Tramacere (2009)}).
We~fit each spectra using three models which are defined as follows:
\begin{enumerate}
\item Power law model, which is defined by $k~E^{\Gamma}$.
It is characterized by photon index $\Gamma$, redshift \emph{z}, and Normalization \emph{k}.

\item Logarithmic parabola model which is defined by
$k~E^{(-\alpha+\beta log(E/E_{1})}$ (e.g., \citep{Massaro (2004),Massaro (2008), Landau (1986)}).
The lowest probed energy $E_{1} = 0.6$ keV is fixed. It is characterized by three
parameters, local spectral index~$\alpha$, curvature parameter $\beta$, and normalization \emph{k}.

\item Broken power law model is defined by $kE^{\Gamma_{1}}$ for $E < E_{\rm break}$ and
$k~E^{\Gamma_{2}}$ otherwise. This model is used to infer the
break energy in those observations which are well fitted by log parabola model. It is characterized by spectral indices $\Gamma_{1}$, $\Gamma_{2}$, break energy $E_{break}$,
redshift, and normalization \emph{k}.

\end{enumerate}


\section{Results}
\subsection{Flux and Spectral Variability}
\begin{figure}[H]
\centering
\includegraphics[width=15cm]{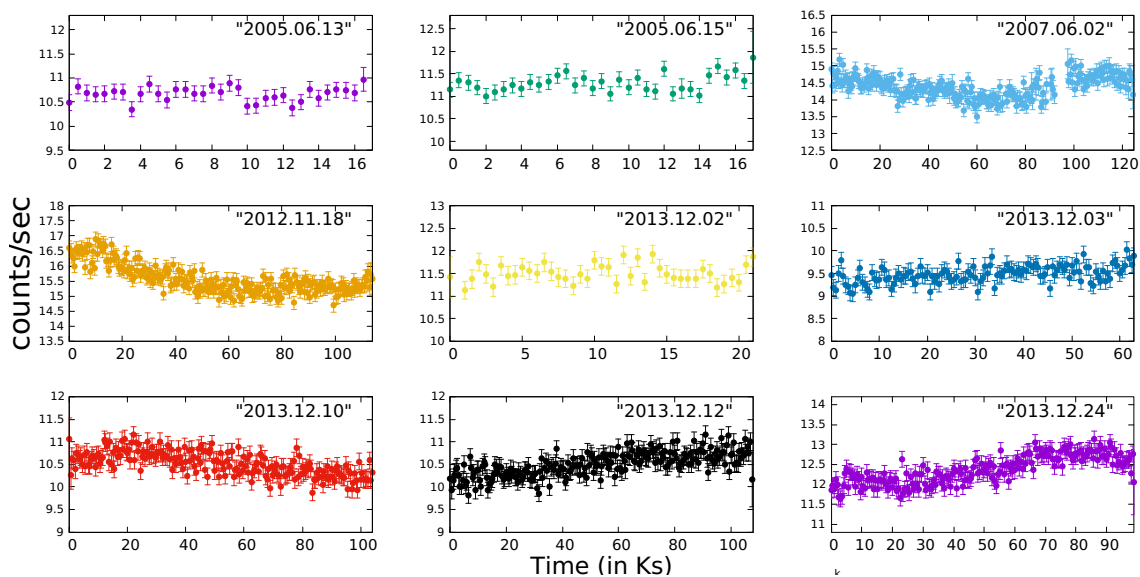}
\caption{ \label{figure:Lightcurve}Light curves of H 2356-309 of XMM-Newton Observations.}
\end{figure}

We analysed nine observations of the blazar H 2356-309 using XMM-Newton satellite data (downloaded from HEASARC Data
archive\footnote{\url{https://heasarc.gsfc.nasa.gov/docs/archive.html}.}). All obtained light curves are shown in
Figure~\ref{figure:Lightcurve}. We~calculated
fractional rms variability amplitude $F_{var}$ in various energy range, i.e., 0.3--10 keV (total), 0.3--2 keV (soft band), and 2--10 keV (Hard band) in order to search for the flux variability in these nine observations. For all LCs X-ray variability parameter $F_{var}$ is presented in Table~\ref{table:X-ray
variability parameter}. It can be seen that the flux variability is observed in five LCs out of total nine LCs. Additionally,
we observed flux variability in soft band (0.3--2 keV) for all of those five LCs. However, in hard band (2--10 keV), out of those five LCs we oserved significant flux variations only in three LCs.

\begin{figure}[H]
\centering
\includegraphics[width=15cm]{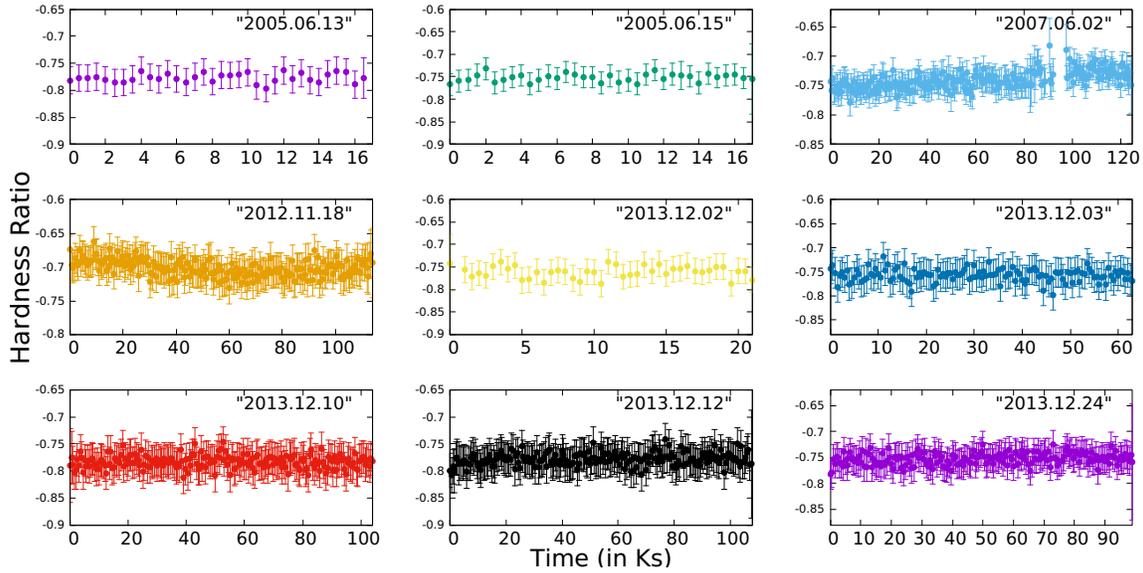}
\caption{Hardness Ratio of H 2356-309.}
\label{figure:Hardness Ratio}
\end{figure}

To search for spectral variability along with flux variability, hardness ratio (HR) is computed. Figure~\ref{figure:Hardness Ratio} shows the~hardness ratio for all of the observations. We did not find any significant spectral change which is obvious from the HR plots. Quantitative variations in HR using a standard $\chi^{2}$ test is obtained, as follows,

\begin{equation}
\chi^{2}=  \sum_i  \frac{(x_{i} - \bar{x})^{2}}{\sigma^{2}_{i}}
\end{equation}
where $x_{i}$ is the HR value, $\sigma_{i}$ is its corresponding error, and $\bar{x}$ is the mean HR value. We only considered a~variation in the HR to be significant if $\chi^{2}$ > $\chi^{2}_{0.90,\:\nu}$ , where $\nu$ is the number of degrees of freedom (DoF) and the significance level is set to 0.90. These results are provided in Table~\ref{table:Temporal variation of Hardness Ratio}, where we see that for each source $\chi^{2}$ < $\chi^{2}_{0.90,\:\nu}$, hence no significant spectral variations are detected in any observation.

\unskip

\begin{table}[H]
\caption{Temporal Variation of hardness ratio (HR) of H-2356 309.}
\label{table:Temporal variation of Hardness Ratio}
\centering
\scalebox{1}[1]{\begin{tabular}{ccccc}
\toprule
\textbf{Obs. Date}	& \multirow{2}{*}{\textbf{DoF}} & \boldmath$ \chi^2$ & \boldmath$ \chi^2_{0.90}$ \\

\textbf{(yyyy-mm-dd)} & &  &  \\
\midrule
2005-06-13 & 33  & 3.63  & 43.745 \\
2005-06-15 & 34  & 4.56  & 44.903 \\
2007-06-02 & 211 & 99.13 & 237.717 \\
2012-11-18 & 225 & 64.10 & 252.578 \\
2013-12-02 & 33  & 9.21  & 43.745 \\
2013-12-03 & 124 & 21.61 & 144.562 \\
2013-12-10 & 207 & 29.30 & 233.466 \\
2013-12-12 & 207 & 30.31 & 233.466 \\
2013-12-24 & 198 & 31.41 & 223.892 \\
\bottomrule
\end{tabular}}\\
\begin{tabular}{ccc}
\multicolumn{1}{c}{\small Bin size = 0.5 ks.}
\end{tabular}


\end{table}

\subsection{Spectral Fitting Results}

\begin{figure}[H]
\centering
\includegraphics[height=4.5 cm, width= 5cm]{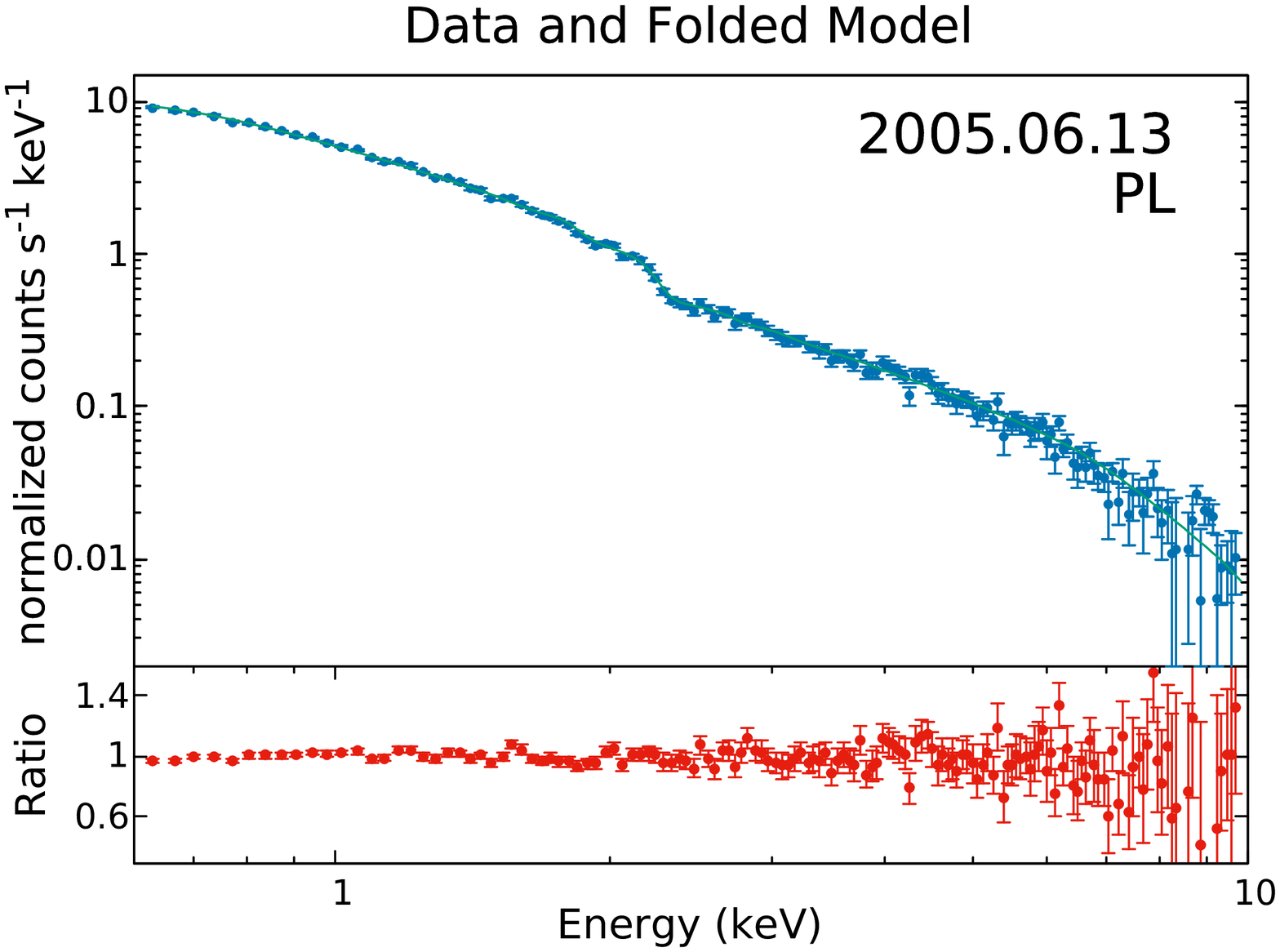}
\includegraphics[height=4.5 cm, width= 5cm]{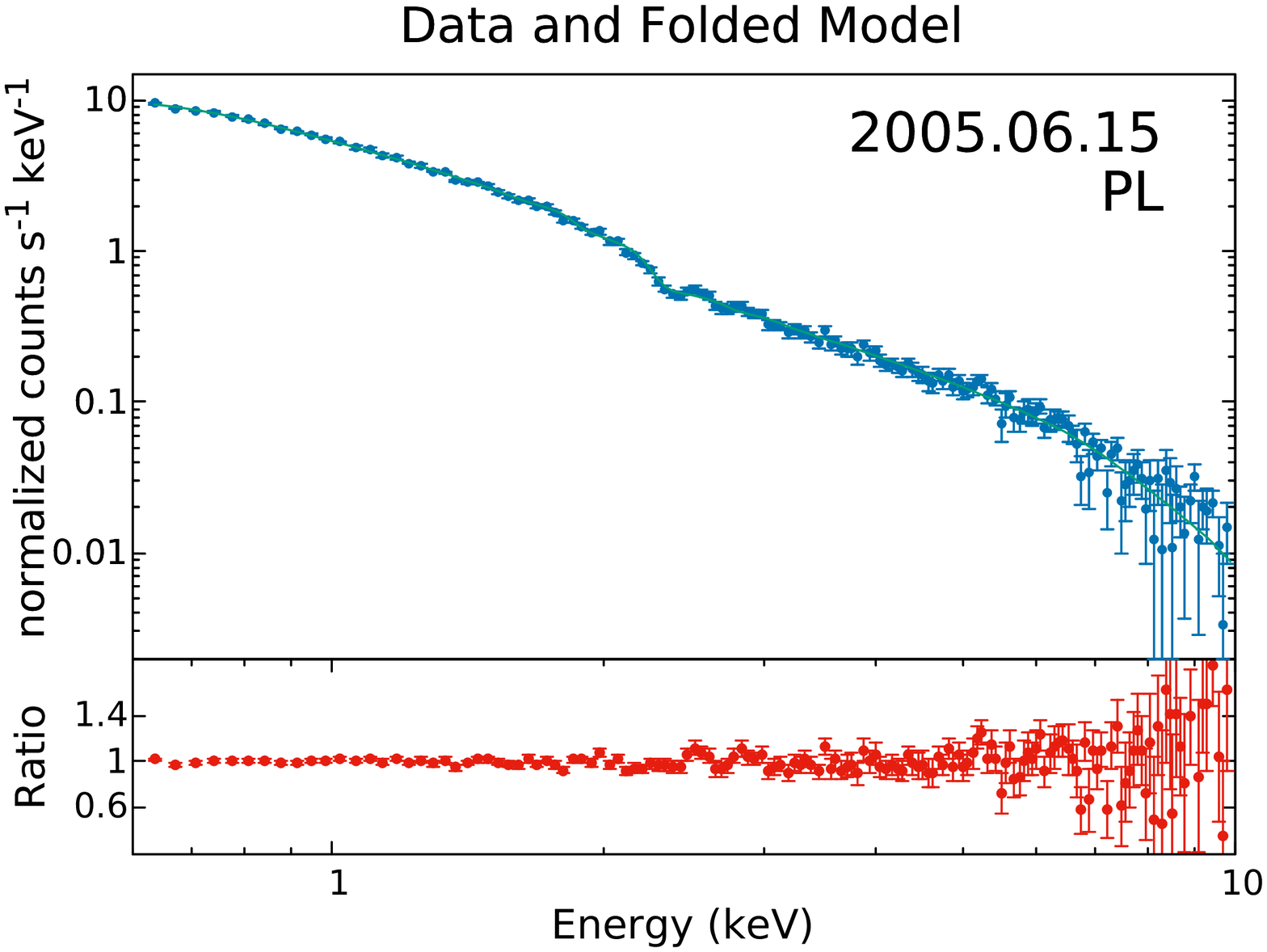}
\includegraphics[height=4.5 cm, width= 5cm]{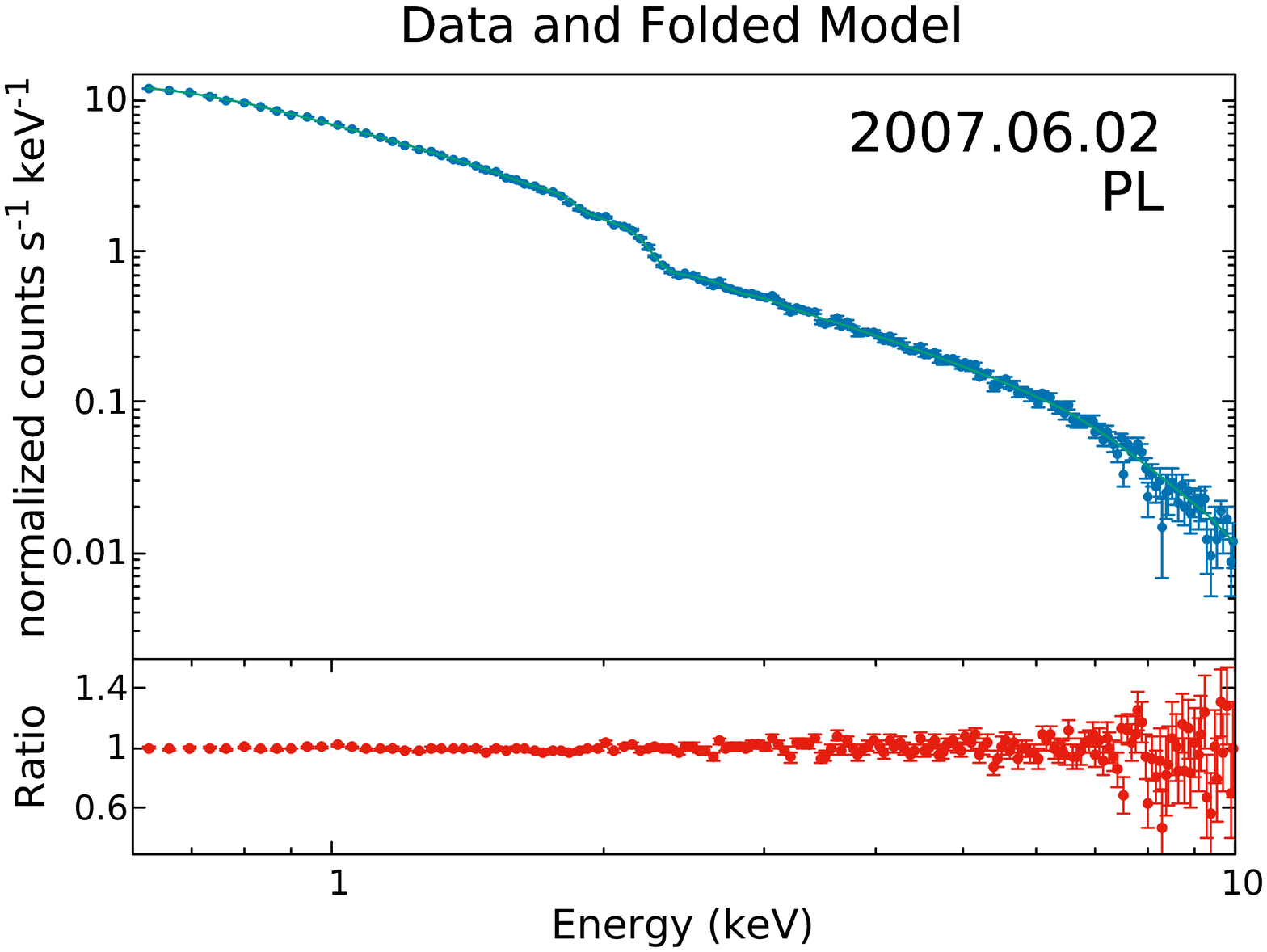}
\includegraphics[height=4.5 cm, width= 5cm]{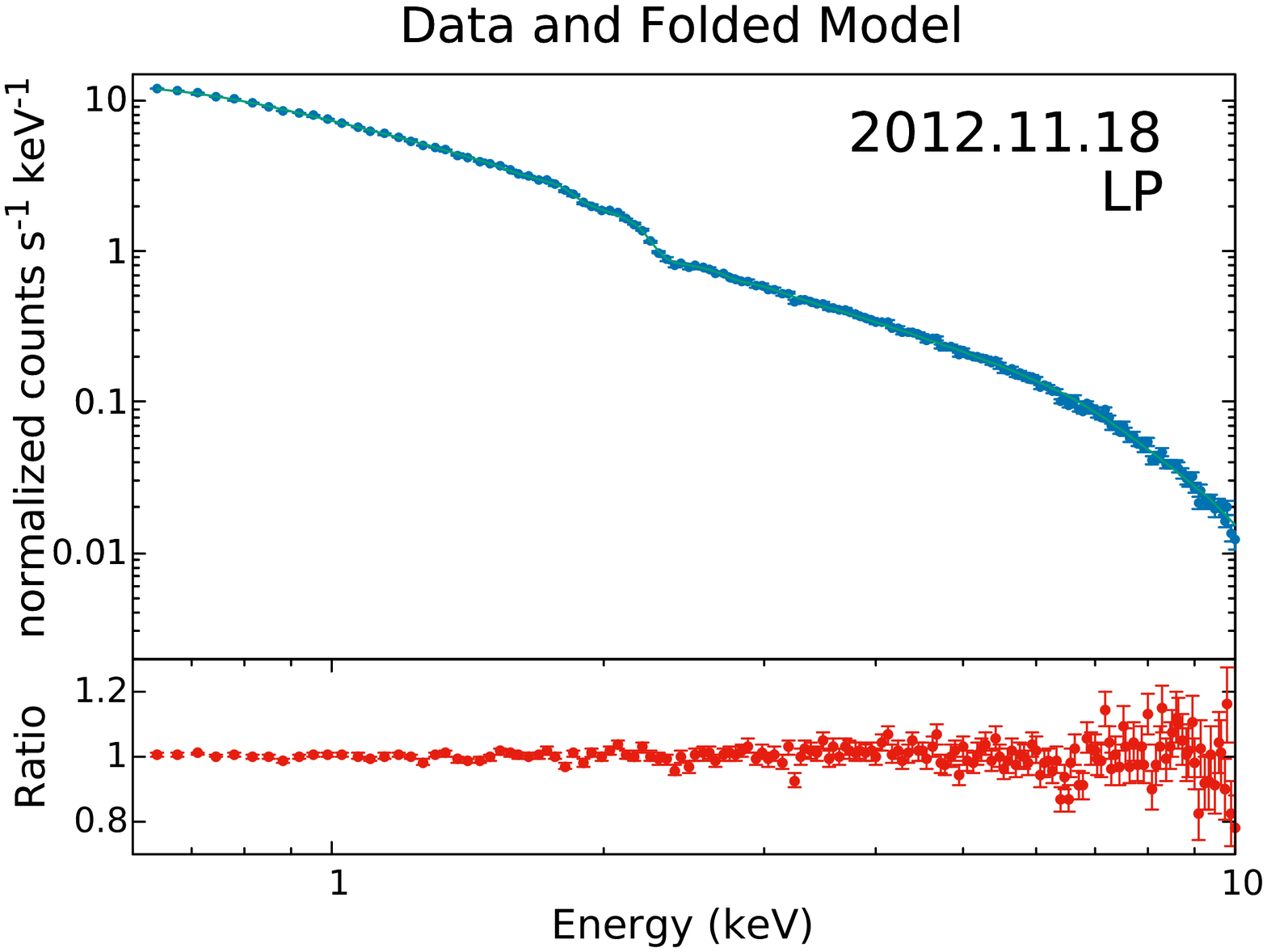}
\includegraphics[height=4.5 cm, width= 5cm]{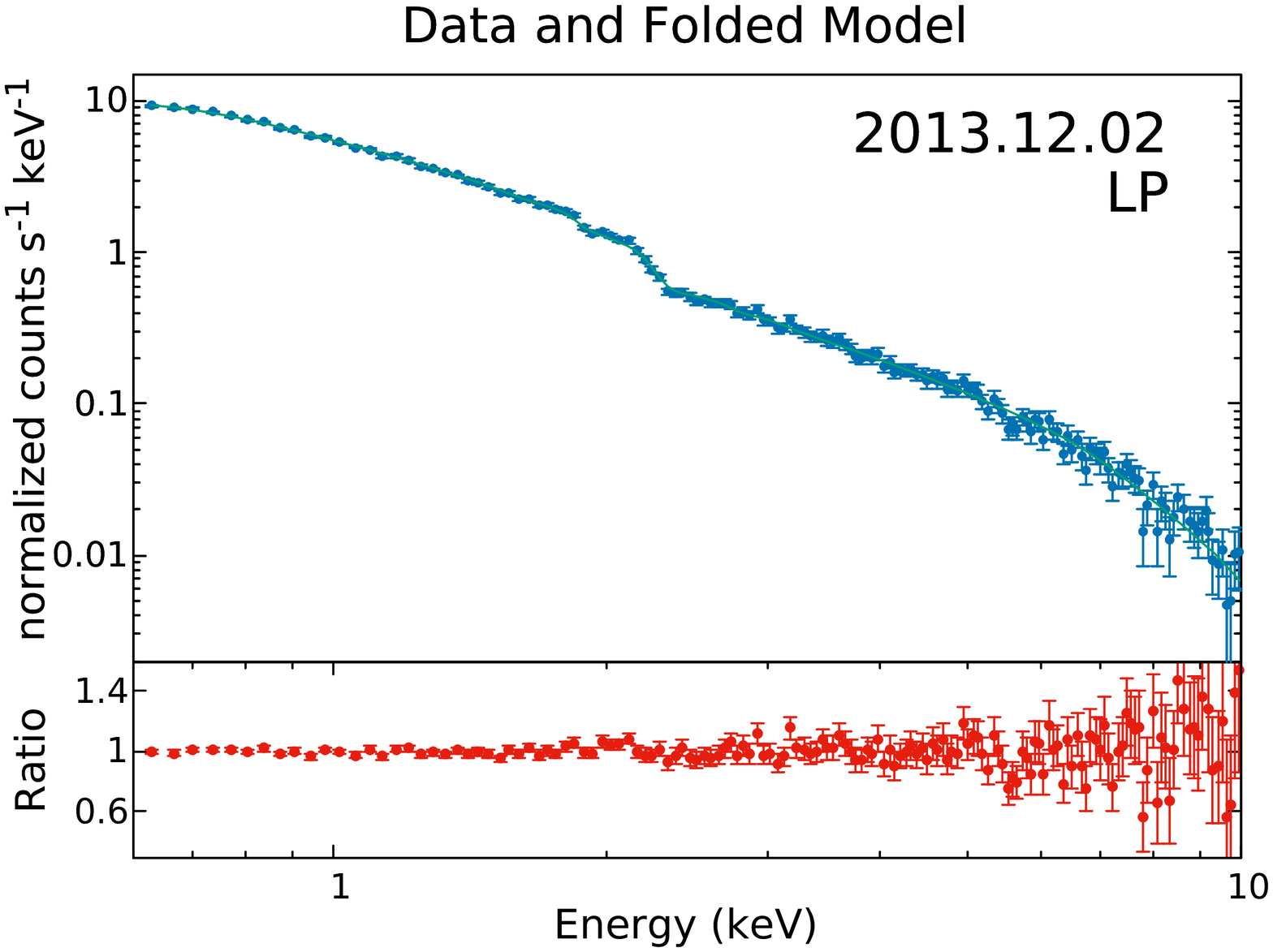}
\includegraphics[height=4.5 cm, width= 5cm]{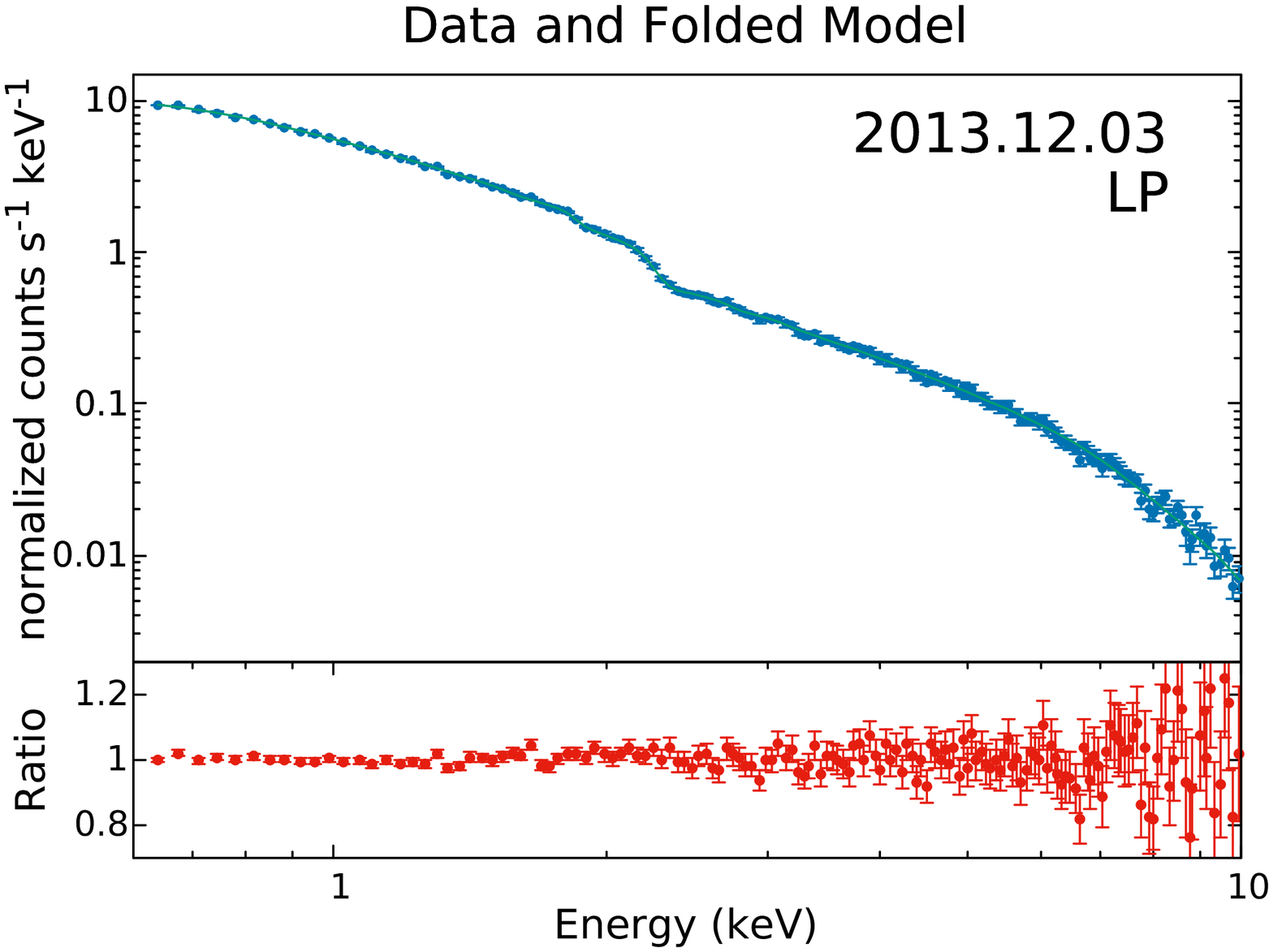}
\includegraphics[height=4.5 cm, width= 5cm]{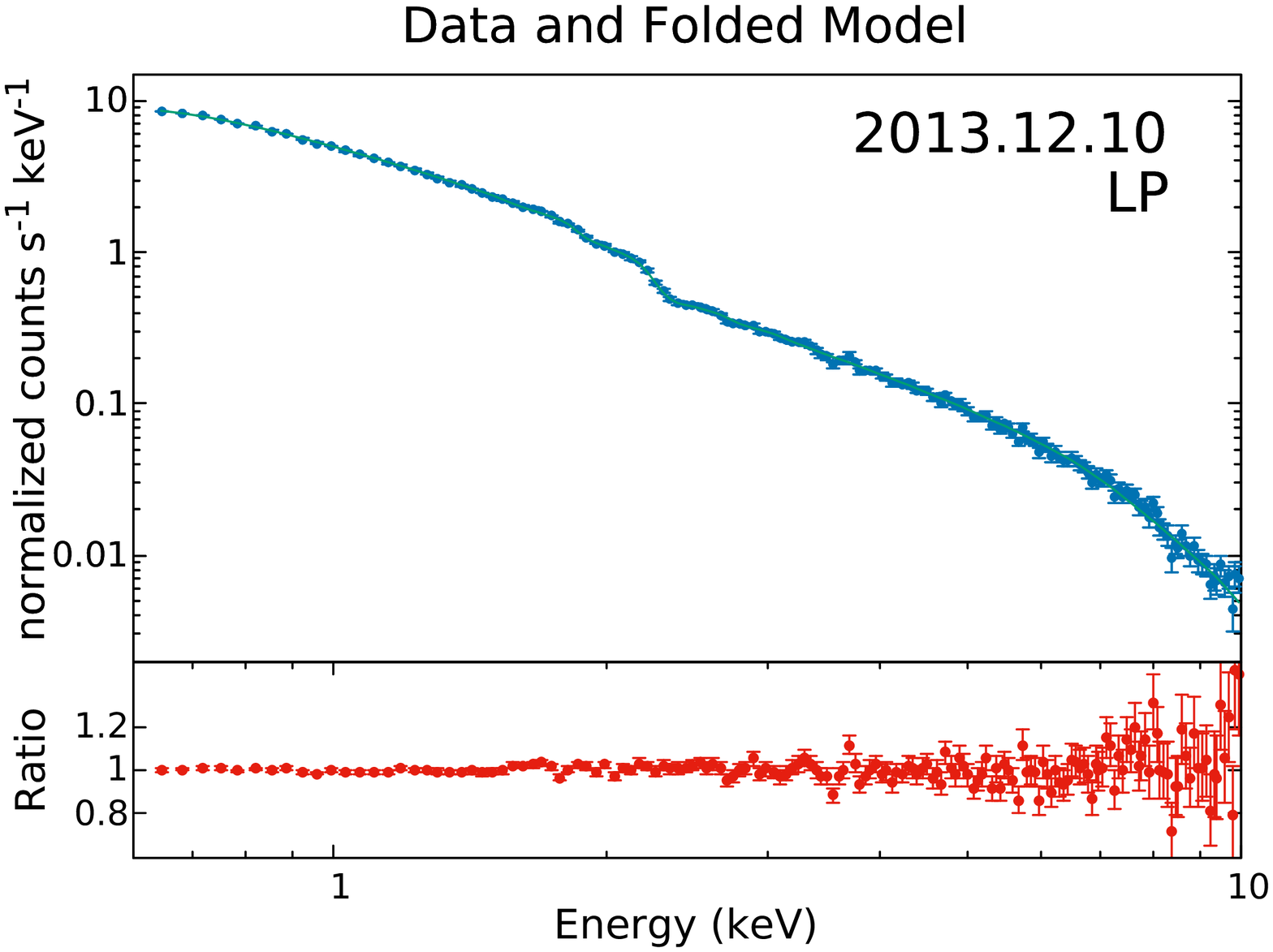}
\includegraphics[height=4.5 cm, width= 5cm]{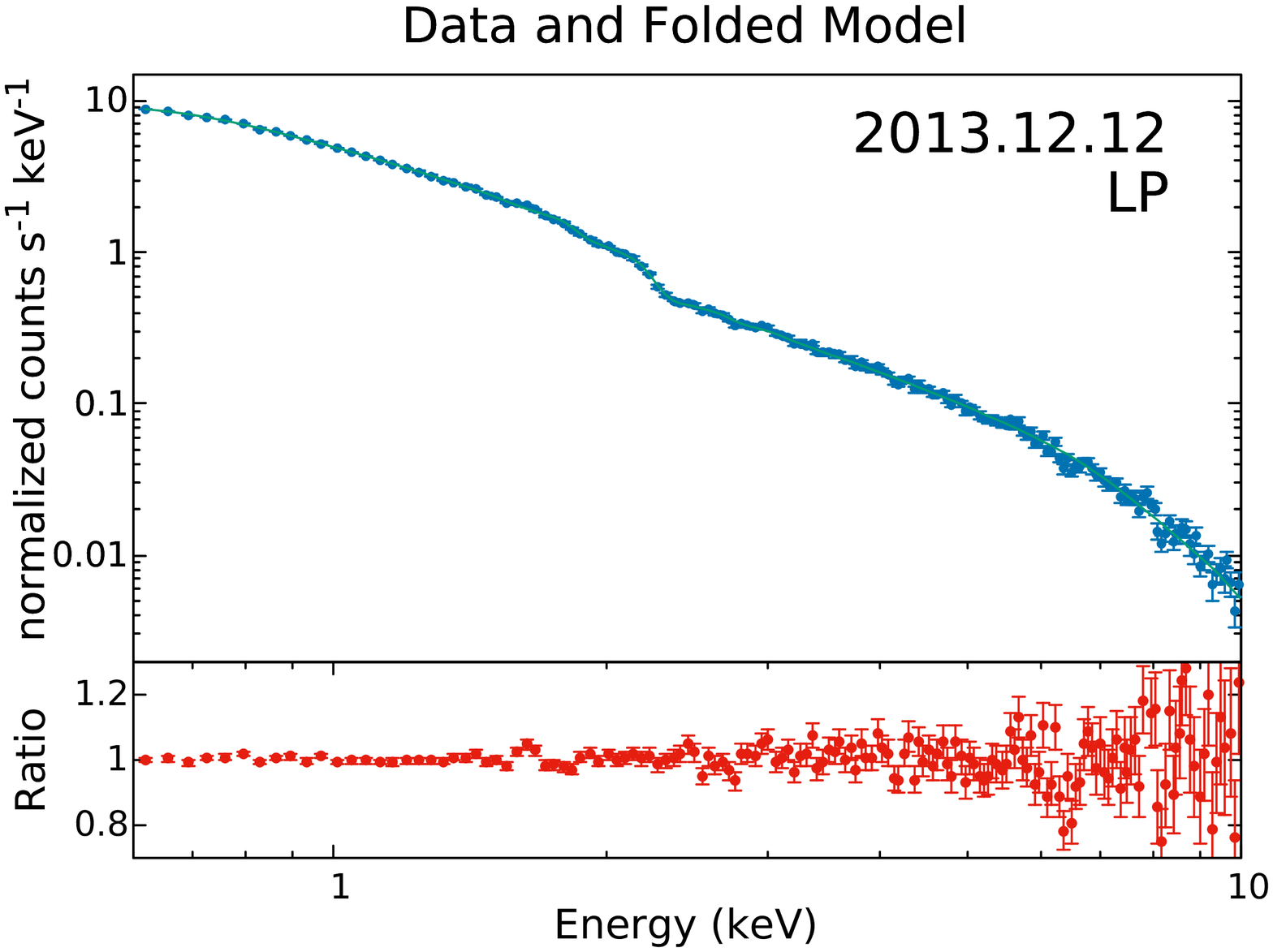}
\includegraphics[height=4.5 cm, width= 5cm]{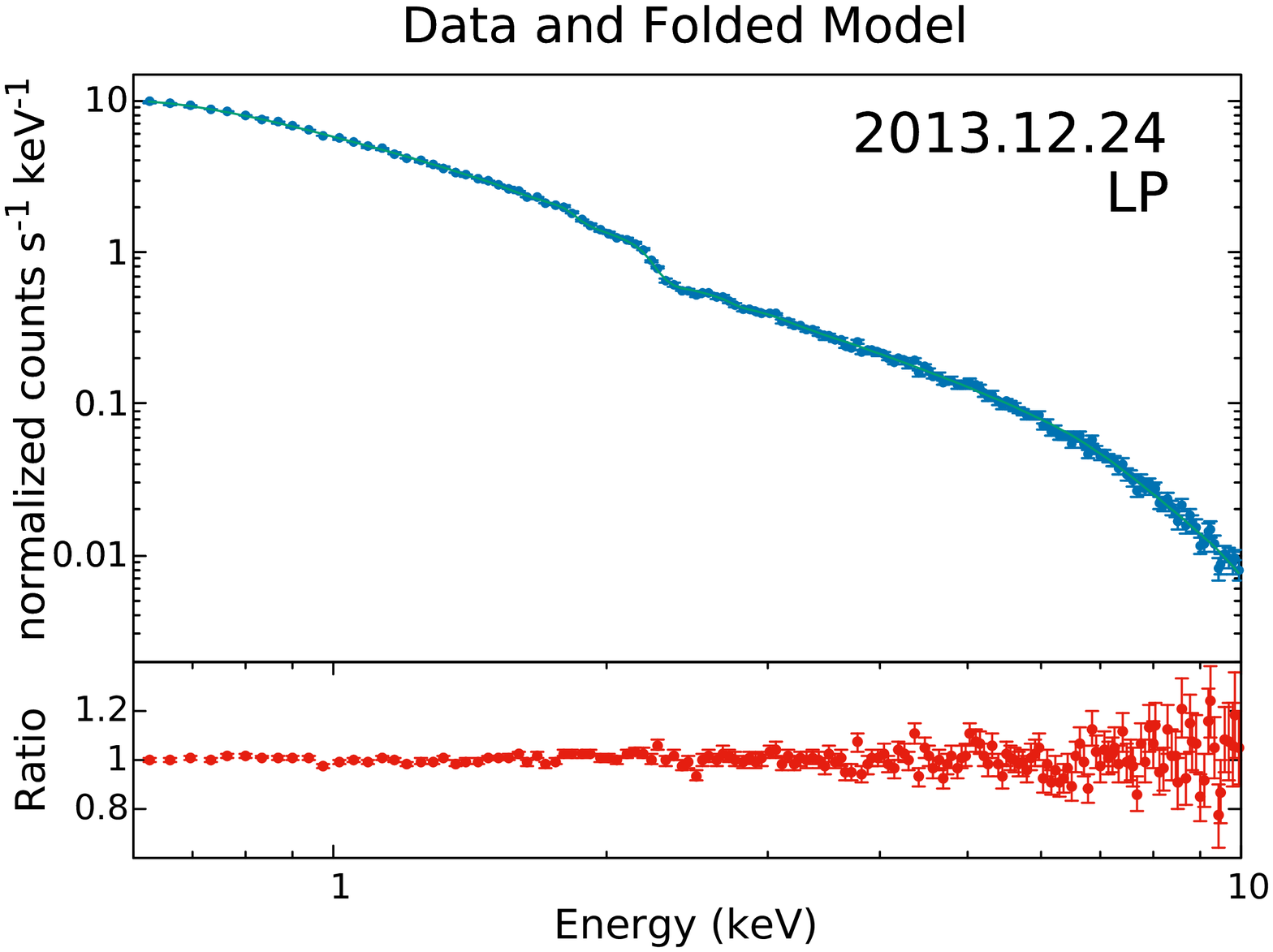}

\caption{The spectra and ratio (data/model) of best fitted model for XMM-Newton observations of \\ H 2356-309.}
\label{figure:Model-fitted spectra}
\end{figure}

We fit all nine observations using power law (PL), log parabolic (LP), and broken power law model (BKN), and the results of spectral
fitting are provided in Table~\ref{table:Spectral Fit Parameters}. To choose the best spectral model between PL and LP model, we performed the F-test using the values
of $\chi^{2}$ and degree of freedom (dof) for both of the models.
LP model provides a better fit than the PL model if the value of F-statistic $>$ 1
and the corresponding null hypothesis probability, \emph{p} $<$ 0.01.
It can be seen from the F-test results that six out of nine observations of the blazar
are well described by the LP model, where $\alpha$ varies between 1.99--2.15 and curvature $\beta$ varies between 0.03--0.18.
Three observations of H 2356-309 are well described by the PL model, as can be seen from the high values (\emph{p} $>$  0.01) of null hypothesis probability. In these observations, $\alpha$ varies between 2.16--2.28.
The observations that are well fitted by log parabolic model are also fitted by broken power law model to constrain the break energy $E_{b}$. It is found that the break energy varies between 1.97--2.31 keV during our observations.
The model-fitted spectra and the data-to-model ratios for each observation are plotted in Figure~\ref{figure:Model-fitted spectra}.

\subsection{Correlation betweeen Various Parameters}

\begin{figure}[H]
\centering
\includegraphics[width=15cm]{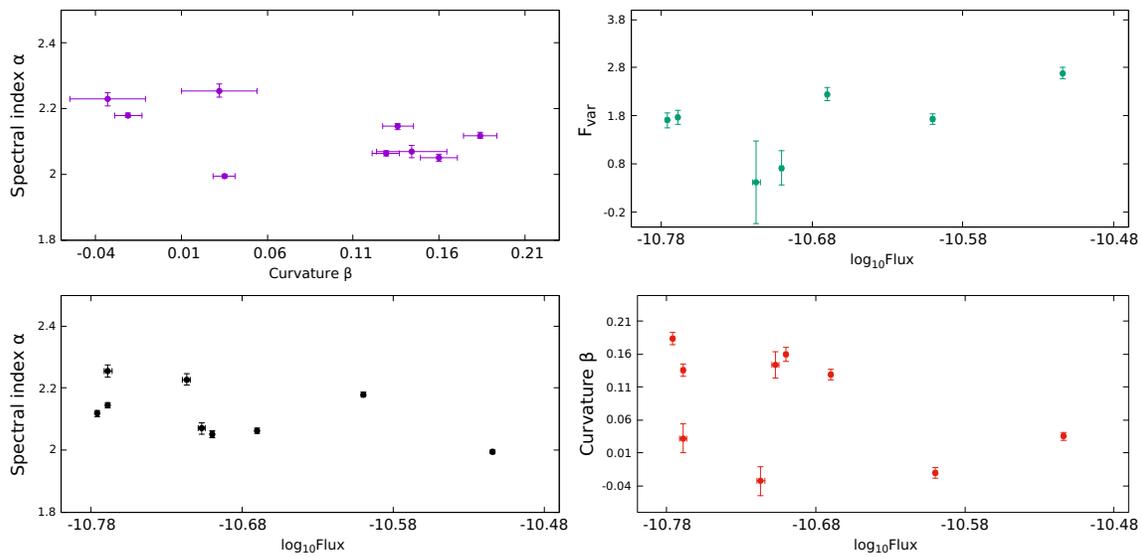}
\caption{Relation between $\alpha$ versus $\beta$; $F_{var}$ versus Flux; $\alpha$ versus Flux and $\beta$ versus Flux.}
\label{figure: Relation between alpha_beta_flux}
\end{figure}

We investigated the correlation between various parameters of LP model i.e., $\alpha$, $\beta$, flux, and $F_{var}$, and all are shown in Figure~\ref{figure: Relation between alpha_beta_flux}.
We derived the Spearman correlation coefficient ($\rho$) between $F_{var}$ versus flux in order to check whether variability amplitude is correlated with flux.
We found \mbox{$\rho$ = 0.5} with null hypothesis probability $p$ = 0.17 between them. Hence, we can conclude that there is no significant correlation between them.
Additionally, we investigated the correlation between spectral index $\alpha$ and flux. From Figure~\ref{figure: Relation between alpha_beta_flux}, it can be seen that there is
an indication of hardening of the spectra as flux increases.
We~found $\rho$ = $-$0.5 with $p$ = 0.13 between $\alpha$ versus flux with \mbox{$\Delta \alpha$ = 0.26} during our observations. Hence,~we~conclude that there is no significant spectral
variations with flux in the blazar, as also suggested by the HR analysis presented above.
We also investigated the correlation between $\beta$ versus flux, which indicates the efficient stochastic acceleration of X-ray emitting electrons~\citep{Massaro (2004)}.
The~correlation value $\rho$ = $-$0.35 with $p$ = 0.39, hence we conclude that there is no significant correlation between them.
Moreover, a positive correlation between $\alpha$ versus $\beta$ is expected within the energy dependent accerelation scenario. We found spearman correlation coefficient \mbox{$\rho$ = $-$0.53} with \mbox{$p$ = 0.14.} Hence, there is a weak negative correlation between them, which indicates that other types of accerelation
mechanisms and cooling processes are at work.

\begin{table}[H]
\caption{Best spectral fit parameters for the Power Law, Log Parabolic and Broken Power Law Model for XMM-Newton Observations of Blazar H 2356-309.}
\label{table:Spectral Fit Parameters}
\centering
\scalebox{.9}[0.9]{\begin{tabular}{ccccccccc}
\toprule
\textbf{Obs. Date}      & \multirow{2}{*}{$\mathbf{\Gamma_{1}}$} & \multirow{2}{*}{\textbf{b} \textbf{or} $\mathbf{\Gamma_{2}}$} & \boldmath{${E_{b}}$ }& $\mathbf{log_{10}Flux}$  & \multirow{2}{*}{\boldmath{$\chi{^{2}_{{Red}}}$}} & \multirow{2}{*}{\textbf{DoF}} & \multirow{2}{*}{\textbf{F-Test}} & \multirow{2}{*}{\textbf{\emph{p}-Value}} \\
\textbf{(yyyy-mm-dd)}& & &  \textbf{(keV)} & \textbf{(ergs/s/cm$^{2}$)} & &  & & \\
\midrule
2005.06.13  & $2.281^{+0.006}_{-0.006}$ & - & - & $-10.767^{+0.002}_{-0.002}$ & 1.043 & 151  & - & - \\
2005.06.15  & $2.200^{+0.006}_{-0.006}$ & - & - & $-10.719^{+0.002}_{-0.002}$ & 0.945 & 153  & - & - \\
2007.06.02  & $2.161^{+0.002}_{-0.002}$ & - & - & $-10.602^{+0.001}_{-0.001}$ & 1.365 & 168  & - & - \\
2012.11.18  & $2.028^{+0.002}_{-0.002}$ & - & - & $-10.512^{+0.001}_{-0.001}$ & 1.492 & 167  & - & - \\
2013.12.02  & $2.197^{+0.006}_{-0.006}$ & - & - & $-10.698^{+0.002}_{-0.002}$ & 1.154 & 152  & - & - \\
2013.12.03  & $2.199^{+0.003}_{-0.003}$ & - & - & $-10.691^{+0.001}_{-0.001}$ & 2.225 & 160  & - & - \\
2013.12.10  & $2.283^{+0.003}_{-0.003}$ & - & - & $-10.766^{+0.001}_{-0.001}$ & 3.348 & 162  & - & - \\
2013.12.12  & $2.266^{+0.003}_{-0.003}$ & - & - & $-10.762^{+0.001}_{-0.001}$ & 2.848 & 163  & - & - \\
2013.12.24  & $2.182^{+0.002}_{-0.002}$ & - & - & $-10.664^{+0.001}_{-0.001}$ & 2.813 & 164  & - & - \\
\midrule
&&\textbf{b} &&&&&& \\
\midrule
2005.06.13  & $2.254^{+0.020}_{-0.020}$ & $0.032^{+0.022}_{-0.022}$ & - & $-10.769^{+0.002}_{-0.002}$ & 1.036 & 150 & 1.013 & 0.316 \\
2005.06.15  & $2.228^{+0.019}_{-0.019}$ & $-0.033^{+0.022}_{-0.022}$ & - & $-10.717^{+0.003}_{-0.003}$ & 0.936 & 152 & 1.461 & 0.229 \\
2007.06.02  & $2.179^{+0.007}_{-0.007}$ & $-0.021^{+0.008}_{-0.008}$ & - & $-10.600^{+0.001}_{-0.001}$ & 1.328 & 167 & 4.653 & 0.032 \\
2012.11.18 {*}   & $1.994^{+0.006}_{-0.006}$ & $0.035^{+0.006}_{-0.006}$ & - & $-10.514^{+0.001}_{-0.001}$ & 1.316 & 166 & 22.201 & $5.167~\times~10^{-6}$ \\
2013.12.02 {*}   & $2.070^{+0.019}_{-0.019}$ & $0.144^{+0.020}_{-0.020}$ & - & $-10.707^{+0.002}_{-0.002}$ & 0.815 & 151 & 62.809 & $4.645~\times~10^{-13}$ \\
2013.12.03 {*}   & $2.051^{+0.010}_{-0.010}$ & $0.160^{+0.011}_{-0.011}$ & - & $-10.700^{+0.001}_{-0.001}$ & 0.834 & 159 & 265.191 & $1.049~\times~10^{-35}$ \\
2013.12.10  {*}   & $2.118^{+0.009}_{-0.009}$ & $0.184^{+0.010}_{-0.010}$ & - & $-10.776^{+0.001}_{-0.001}$ & 1.048 & 161 & 353.34 & $1.871~\times~10^{-42}$ \\
2013.12.12 {*}   & $2.145^{+0.008}_{-0.008}$ & $0.136^{+0.009}_{-0.009}$ & - & $-10.769^{+0.001}_{-0.001}$ & 1.372 & 162 & 174.28 & $1.757~\times~10^{-27}$ \\
2013.12.24 {*}   & $2.063^{+0.008}_{-0.008}$ & $0.129^{+0.008}_{-0.008}$ & - & $-10.670^{+0.001}_{-0.001}$ & 1.225 & 163 & 211.301 & $3.112~\times~10^{-31}$ \\
\midrule
&&$\mathbf{\Gamma_{2}}$&&&&&& \\
\midrule
2012.11.18  & $2.014^{+0.004}_{-0.004}$ & $2.056^{+0.008}_{-0.008}$ & $2.314^{+0.327}_{-0.327}$ & $-10.514^{+0.001}_{-0.001}$ & 1.326 & 165 & - & - \\
2013.12.02  & $2.143^{+0.011}_{-0.011}$ & $2.311^{+0.023}_{-0.023}$ & $2.056^{+0.197}_{-0.197}$ & $-10.707^{+0.002}_{-0.002}$ & 0.804 & 150 & - & - \\
2013.12.03  & $2.135^{+0.006}_{-0.006}$ & $2.322^{+0.010}_{-0.010}$ & $2.087^{+0.103}_{-0.103}$ & $-10.698^{+0.001}_{-0.001}$ & 0.786 & 158 & - & - \\
2013.12.10  & $2.210^{+0.005}_{-0.005}$ & $2.414^{+0.010}_{-0.010}$ & $1.967^{+0.074}_{-0.074}$ & $-10.775^{+0.001}_{-0.001}$ & 1.026 & 160 & - & - \\
2013.12.12  & $2.217^{+0.005}_{-0.005}$ & $2.372^{+0.011}_{-0.011}$ & $2.098^{+0.107}_{-0.107}$ & $-10.769^{+0.001}_{-0.001}$ & 1.484 & 161 & - & - \\
2013.12.24  & $2.130^{+0.005}_{-0.005}$ & $2.286^{+0.010}_{-0.010}$ & $2.137^{+0.100}_{-0.100}$ & $-10.670^{+0.001}_{-0.001}$ & 1.174 & 162 & - & - \\

\bottomrule
\end{tabular}}\\
\begin{tabular}{ccc}
\multicolumn{1}{p{\textwidth-.68in}}{\small $\Gamma_{1}$: Low energy spectral index;  $\Gamma_{2}$: High energy spectral index; b: curvature; $E_{b}$: Break Energy; $\chi^{2}_{Red}$: Reduced $\chi^{2}$;  DoF: degree of freedom; * indicates the observations that are well fitted by the LP model.}
\end{tabular}

\end{table}
\unskip

\section{Discussion}

The study of X-ray flux variability on intra-day timescales is an important tool to understand the emission mechanism in blazars.
It can be used to estimate the size and constrain the location and structure of a dominant emitting region (e.g., \citep{Ciprini (2003)}). In blazars, most intrinsic flux variability across the EM bands can be explained by standard relativistic jet-based models (e.g., \citep{Marscher (1985),Calafut (2015)}), as it is closely aligned to our line of sight, while accretion-disk-based models are not very important for such sources, particularly when they are in high state. Accretion disk contribution is noticable in blazars, when they are observed in a low flux state \citep{Ciprini (2003),Calafut (2015)}.

For HSP blazars, first hump of the SED peaks in UV/X-rays and, hence, they are brightest in X-ray bands. They show strong variability in this band, along with the large amplitude variations.
In this work, we presented the timing and spectral analysis of nine observations of the HSP blazar H 2356-309 performed during 2005--2013.
We used the fractional variability method to examine the
intra-day variability in all of the light curves. For the HSP blazar H 2356-309 studied here, we found significant flux variability
in five of the light curves out of total nine observations. Amplitude variations are between 1.71--2.69 \% for these observations.

As we already mentioned that the accepted model for the strong variability in blazars on long term timescales involve shocks propagating down relativistic jets
pointing close to our line of sight (i.e., \citep{Marscher (1985), Wagner (1995)}). The shorter variations or variations on IDV timescales can be mostly explained in terms of irregularities in the jet flows; turbulence behind the jet \citep{Calafut (2015)}; jets-in-a-jet model (i.e., \citep{Giannios (2009)}); magnetic reconnections
taking place in fast-moving emitting regions within jets that accelerate the particles to high energies \citep{Paliya (2015)}.
The shortest variability in the X-ray band (3--79 keV) is shown by the TeV blazar Mrk 421 with doubling time of $\sim$14 minutes during its April 2013 flare \citep{Paliya (2015)}.

We also examined the X-ray spectral variability of blazar H 2356-309 using the HR analysis. From the $\chi^{2}$ values in all of the observations,
we found that there is no significant spectral variations in any of the observations. The fractional variability amplitude of TeV HSPs are studied in detail in literature and it is found that it ranges from few per cent to around 50\% during flaring events (e.g., \citep{Pandey (2018),Pandey (2017),Aggrawal (2018)}). Ref. \citep{Pandey (2018)} also studied the blazar H 2356-309 in 2018 using the NuSTAR observations, but did not find any significant flux or spectral~variability.

It is well known that the spectra of high energy peaked blazars is intrinsically curved and, hence, log parabolic model better describes their spectra. The curved spectra are thought to originate from an
energy dependent particle acceleration process (\citep{Massaro (2004), Massaro (2008), Tramacere (2007a), Gaur (2017)}).
The curvature of the X-ray spectra can be either convex or concave. Convex curvature of the X-ray spectra, is likely to be caused by a single accelerated particle
distribution (e.g., \citep{Massaro (2004)}). The concave X-ray spectra may be a consequence of the spectral upturn at the interaction of the high-energy tail of the
synchrotron emission and the low-energy part of the inverse Compton emission (e.g., \citep{Gaur (2018),Wierzcholska (2016)}).
In order to study the X-ray spectral shape of blazar H 2356-309, we fit all of our observations using the power law and log parabolic model. We~found that only three of our
observations are well fitted by the power law model. Six of the observations showed spectral curvature and are well fitted by log parabolic model. Their photon indices varies between 1.99--2.15 and curvature values varies between 0.03--0.18.

In the energy dependent acceleration mechanism, a positive correlation is expected between $\alpha$ and $\beta$ (\citep{Massaro (2004),Massaro (2008)}). However, in our observations
of the blazar H 2356-309, we found a weak negative correlation between them. Additionally, we searched for correlation between $\alpha$ versus flux in order to search for
spectral variability but no significant variation is found between them. These results are in agreement with the studies of HR analysis.
No correlation is found between $\beta$ versus flux and amplitude of variability versus flux.

\section{Conclusions}
We studied nine XMM--Newton observations of the HSP blazar H 2356-309, which are available in its public
archive. Our~conclusions are summarized, as follows:

\begin{enumerate}
\item The fractional variability analysis shows that there is moderate amplitude IDV for five of the total nine observations. Flux variability is observed in soft band (0.3--2 keV) in all of these 5~LCs, but~only three light curves showed flux variability in hard band (2--10 keV).
\item The variability amplitude is lower in the soft bands than in the hard bands that can be incorporated in energy dependent synchrotron model.
We also estimated the IDV timescale for all
nine observation IDs in 0.3--2 keV (soft band) and 2--10 keV (hard band), but did not find any significant variability timescales.
\item HR analysis of all the nine observations shows that there is no significant spectral variability associated with the flux variability.
\item We fit all the spectra with power law and log parabolic model. Six of the observations are well fitted by log parabolic model indicating curvature
ranging between 0.03--0.18 and alpha varies between 1.99--2.15. Remaining three observations are well fitted by the power law model. The~break energy is constrained to be $\sim$2.1 keV.
\item We searched for the correlations between $\alpha$ versus $\beta$, but found very weak negative correlation between them. We did not find any significant correlation
between $\beta$ versus flux. No significant correlation between $\alpha$ versus flux is found, which again indicates no significant spectral variations in this source.
\end{enumerate}

\authorcontributions{K.A.W. analyzed the X-ray data, carried out timing and spectral analysis and wrote the initial manuscript. H.G. conceived the idea and finalize the manuscript.  All authors have read and agreed to the published version of the manuscript.
}

\funding{This research is based on observations obtained with XMM-Newton, an ESA science mission with instruments and contributions directly
funded by ESA member states and NASA.  
}
\acknowledgments{Authors would like to thank the referees for their useful comments.
Authors acknowledge the financial support from the Department of Science and Technology, India, through INSPIRE
faculty award IFA17-PH197 at ARIES, Nainital.
}
\conflictsofinterest{The authors declare no conflict of interest.}


\reftitle{References}

\end{document}